\documentclass[aps,prl,twocolumn,floatfix,english,showpacs,10pt,superscriptaddress]{revtex4-2}%
\usepackage{graphicx}
\usepackage{float} 
\usepackage{booktabs}
\usepackage{amsmath}
\usepackage{physics}
\usepackage{amssymb}
\usepackage{mathrsfs}
\usepackage{colordvi}
\usepackage{verbatim}
\usepackage{xcolor}
\usepackage{multirow}
\usepackage{mathrsfs}
\usepackage{epsfig}
\usepackage{lipsum}
\usepackage{enumitem}
\usepackage{amsfonts}
\usepackage[unicode=true, breaklinks=false, pdfborder={0 0 1}, backref=false,
colorlinks=true, linkcolor=blue, urlcolor=blue, citecolor=blue]{hyperref}%
\providecommand{\U}[1]{\protect\rule{.1in}{.1in}}

\setcitestyle{numbers,square}
\begin{document}
\title{Scale-Free Response with Directional Amplification in Critical Non-Hermitian Systems}
\author{Kunling Zhou}


\affiliation{School of Physics, Huazhong University of Science and Technology, Wuhan 430074, P. R. China}

\author{Zihe Yang}


\affiliation{School of Physics, Huazhong University of Science and Technology, Wuhan 430074, P. R. China}

\author{Bowen Zeng}
\email[]{zengbowen@csust.edu.cn}
\affiliation{Hunan Provincial Key Laboratory of Flexible Electronic Materials Genome Engineering,
School of Physics and Electronic Sciences, Changsha University of Science and Technology, Changsha 410114, P. R. China}

\author{Yong Hu}
\email[]{huyong@hust.edu.cn}
\affiliation{School of Physics, Huazhong University of Science and Technology, Wuhan 430074, P. R. China}

\makeatletter
\newcommand{\rmnum}[1]{\romannumeral #1}
\newcommand{\Rmnum}[1]{\expandafter\@slowromancap\romannumeral #1@}
\makeatother

\begin{abstract}
The non-Hermitian skin effect can lead to directional amplification of response, with the associated end-to-end Green's function generally exhibiting size dependence. Any deviation in length or local disorder can drastically alter the amplification factor, limiting the practicality of the response in implementations. In this work, we identify a new type of scale‑free, topological, and directionally amplified response in a Hatano–Nelson model under perturbed open boundary conditions.  The scale-free response can be attributed to the first order boundary effect and characterized by a winding number defined on a continuous generalization of the finite-size Brillouin zone—a concept introduced in this work. Such scale-free behavior endows the end‑to‑end Green’s function with significant robustness and making it promising for practical applications.
\end{abstract}

\maketitle

~\textit{Introduction—}The response of a system to external perturbations holds fundamental significance, both as a window into underlying physics and as a key to practical applications~\cite{mahan2013many,chaikin1995principles}. In the context of detecting weak signals in complex environments, a response endowed with three particular features is especially desirable:  (i) topological protection, which ensures robustness against disorder~\cite{PhysRevLett.45.494,PhysRevB.23.5632,PhysRevLett.49.405,RevModPhys.82.1959,wanjura2020topological,PhysRevLett.122.143901,PhysRevA.103.033513}; (ii) directional amplification, enabling  enhancement of weak signals and  suppression of back-action to the source~\cite{wanjura2020topological,PhysRevB.106.024301,PhysRevX.8.041031,PhysRevLett.122.143901,PhysRevA.103.033513}; and (iii) a scale-free nature, which ensures consistent behavior across different system sizes and thus allows small-scale systems to faithfully represent their large-scale counterparts~\cite{razo2025scale,PhysRevB.109.L140102,HaoWang2025014203}. While each of these features can be realized individually, combining them in a single system remains a significant challenge.

In Hermitian lattice systems,  the response to an external perturbation generally decays spatially, thereby preventing directional amplification~\cite{Wangzhong2021Simpleformulas,zirnstein2021exponentially}. Recently, non-Hermitian systems of general interest~\cite{doi:10.1080/00018732.2021.1876991,Griffiths_Schroeter_2018,Gongzp2018topologicalphases,okuma2020topological,yokomizo2019non,PhysRevLett.116.133903,PhysRevLett.121.086803,RevModPhys.93.015005,PhysRevLett.121.026808,lee2019anatomy,yu2024non,PhysRevLett.124.056802,hu2025acoustic,wu2025hybrid} have emerged as a vibrant research frontier in the study of excitation and response~\cite{PhysRevLett.128.120401,3ht3-ty3h,PhysRevB.109.214311,2025arXiv251220106L,PhysRevB.106.L241112,PhysRevX.8.041031,PhysRevB.110.L140302,li2024observation,PhysRevLett.133.070801,PhysRevLett.132.203801,doppler2016dynamically,tian2023nonreciprocal}. A hallmark of such systems is the non-Hermitian skin effect~\cite{okuma2020topological,lee2019anatomy}, a phenomenon absent in Hermitian systems where all bulk states accumulate at boundaries, enabling exponential amplification along a preferred direction and exponential suppression along the opposite direction~\cite{Wangzhong2021Simpleformulas,wanjura2020topological,PhysRevResearch.5.043073,PhysRevB.107.115412}. Although the unidirectionality of the response is  topologically protected, the amplification is sensitive to the variations in system size and parameters, limiting the practicality of the response in realistic implementations~\cite{Wangzhong2021Simpleformulas,PhysRevB.105.045122}. Recently, scale-free characteristics have been observed in  systems with critical non-Hermitian skin effect, boundary perturbation, and local non-Hermiticity, where for specified eigenstates, the boundary accumulation is scale-independent~\cite{li2020critical,PhysRevB.104.165117,PhysRevLett.127.116801,li2021impurity,2026arXiv260322746L,PhysRevB.107.134121,PhysRevB.108.L161409,PhysRevB.109.035119,PhysRevB.109.L140102,HaoWang2025014203}. However, extending this scale-free to dynamical responses-while simultaneously preserving topological protection and directional amplification—has remained elusive. 

In this work, we investigate  the response property of a  non-Hermitian Hatano-Nelson model  with a slightly perturbed open boundary condition (pOBC) by calculating its Green's function. Unlike previous results where the end-to-end Green's function exhibits exponential amplification or suppression with system size~\cite{Wangzhong2021Simpleformulas,wanjura2020topological,PhysRevB.105.045122}, we find that it may remain unchanged as the system size varies—a behavior we term scale-free response. Although the spectrum of this system is discrete and size-dependent, we can propose a continuum limit approach for the finite-size generalized Brillouin zone (fGBZ)~\cite{PhysRevLett.127.116801}. The scale-free response is topologically protected by a winding number defined on this continuous fGBZ with respect to the driving frequency. In topologically trivial and nontrivial regions, the end‑to‑end Green's function either (i) shows scale‑free amplification and exponential suppression in the opposite direction, or (ii) shows scale‑free suppression and exponential amplification in the opposite direction. Furthermore, this scale-free amplification exhibits robustness to variations in system parameters. These properties offer a promising route toward realizing robust and efficient amplifiers.

\begin{figure*}[t]
    \centering
    \includegraphics[width=0.85\linewidth]{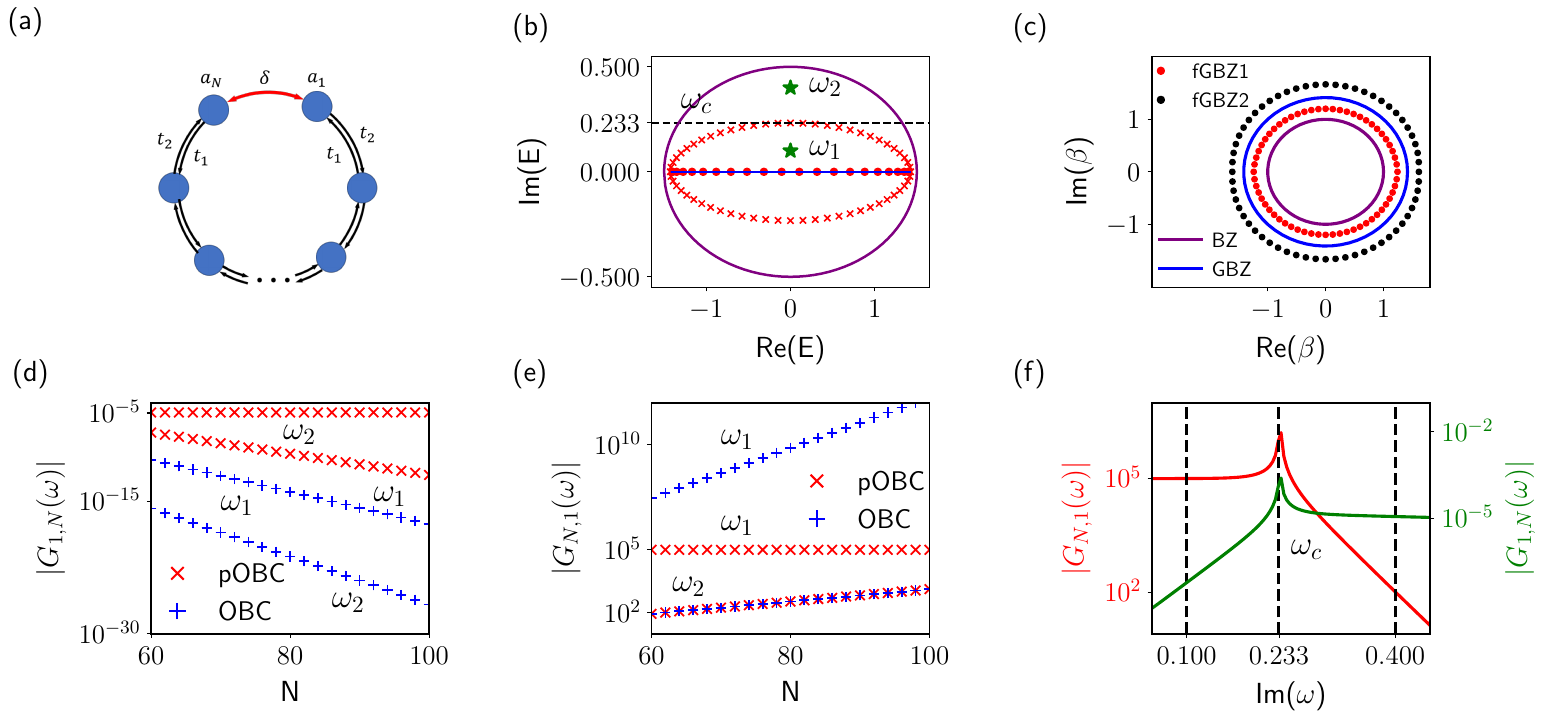}
    \caption{(a) Schematic of the Hatano-Nelson model with nonreciprocal couplings $t_1=0.5$ and $t_2=1$, where the two ends are connected via a perturbation $\delta=0.00001$ .
    (b) Energy spectrum of the model. The spectra under thermodynamic limit for PBC and OBC are denoted by purple curve, blue curve, while red dots and red ``x"  correspond to the pOBC spectra with $N=20$ and $N=60$. The critical frequency $\omega_c$ lies between the external and internal domains of pOBC spectrum. The frequency $\omega_1=0.1i$ and $\omega_2=0.4i$. (c) The BZ, GBZ, fGBZ1, fGBZ2 for the model. Dependence of the Green's functions (d) $G_{1,N}(\omega)$ and (e) $G_{N,1}(\omega)$  for $\omega_1$ and $\omega_2$ with the lattice size $N$. All end-to-end Green's functions are size-dependent, except $G_{1,N}(\omega_2)$ and $G_{N,1}(\omega_1)$ under pOBC, which remain scale invariant.
    (f) Frequency dependence of $G_{N,1}(\omega)$ and $G_{1,N}(\omega)$，for $N=60$. When the frequency crosses $\omega_c$, the Green's functions under pOBC undergo a transition.}
    \label{Fig1}
\end{figure*}

~\textit{The scale-free response—}We start with a finite-sized Hatano-Nelson model with pOBC, where a small coupling $\delta$ is introduced between the two ends, as illustrated in Fig.~\ref{Fig1}(a).
The Hamiltonian is governed by 
\begin{align}     
\label{eq-secII-Hatano}
H=\sum_{n=1}^{N-1}(t_1a_n^{\dagger}a_{n+1}+t_2a_{n+1}^{\dagger} a_n) + \delta a_1^{\dagger}a_N+\delta a_N^{\dagger}a_1.
\end{align}
Without loss of generality, here we assume that the non-reciprocal hoppings satisfy $t_2 > t_1$.

When $\delta=0$, the system reduces to open boundary conditions (OBC). In the thermodynamic limit, according to non-Bloch band theory~\cite{yokomizo2019non,PhysRevLett.125.126402}, the generalized Brillouin zone (GBZ) is calculated by the bulk equation 
\begin{equation}
\label{eq-secII-Hatanobulk}
    E(\beta)= t_1\beta+t_2/\beta,
\end{equation}
with equal modulus $|\beta_1|=|\beta_2|=\sqrt{t_2/t_1}$. The GBZ and corresponding OBC spectrum are shown in Fig.~\ref{Fig1}(c) and (b). In the above figures, we also plot the  Brillouin zone (BZ) and PBC spectrum as a reference. Unlike PBC and OBC spectra, which are both derived in the infinite-length limit, the numerical calculated spectrum under pOBC strongly depends on the size [see Fig.~\ref{Fig1}(b)]. When $N=20$ and $\delta=0.00001$, this discrete spectrum lies entirely within the OBC spectrum. However, when $N=60$, this discrete spectrum appears to form a loop similar to the PBC spectrum. Indeed, as the system size increases, the spectrum approaches the PBC spectrum (not shown), which is known as non-Hermitian critical phenomena~\cite{li2020critical,PhysRevLett.127.116801}. According to Eq.~\eqref{eq-secII-Hatanobulk}, the two roots corresponding to this spectrum are not equal, forming two rings divided by GBZ in Fig.~\ref{Fig1}(c), which is known as  fGBZ~\cite{PhysRevLett.127.116801}. Assuming $\abs{\beta_1} < \abs{\beta_2}$,  the set of $\beta_1$ (red dots) is denoted as fGBZ1, and that of $\beta_2$ (black dots) as fGBZ2.

We now turn to the dynamical excitation and response by considering frequency-domain Green’s function  $G(\omega)=1/(\omega-H)$. Under OBC, it is demonstrated that the end-to-end Green’s function for excitation frequency $\omega$ has the form of $G_{N,1}(\omega) \propto [\beta_1(\omega)]^L,G_{1,N}(\omega) \propto [\beta_2(\omega)]^{-L}$~\cite{Wangzhong2021Simpleformulas}. Here, $G_{N,1}(\omega)$ denotes the response at site $N$ and the excitation at site $1$ for excitation frequency $\omega$, and $\beta_1(\omega),\beta_2(\omega)$ are two roots according to Eq.~\eqref{eq-secII-Hatanobulk}. These particular forms follow from the fact that the integration path of Green's function is chosen as the GBZ~\cite{Wangzhong2021Simpleformulas,PhysRevB.107.115412}, so that one root is inside the contour and the other is outside~\cite{PhysRevLett.125.126402}. Indeed, as shown in Fig.~\ref{Fig1}(d) and (e), the Green's function on a logarithmic scale scales linearly with the size.

Under pOBC, $G_{1,N}(\omega_1)$ and $G_{N,1}(\omega_2)$ behave similarly to that under OBC. However, unlike OBC, $G_{1,N}(\omega_2)$ and $G_{N,1}(\omega_1)$ exhibit suppression and amplification that do not depend on the size, termed as scale-free amplification and suppression, as seen in Fig.~\ref{Fig1}(d) and (e). Compared with the winding number under OBC~\cite{PhysRevLett.125.126402},
\begin{align}
    \label{eq-secII-winding}
    W_{\text{GBZ},\omega} = \frac{1}{2\pi} \oint_{\text{GBZ}} \frac{d}{d\beta} \arg(E(\beta)-\omega)  d\beta,
\end{align}
here the fGBZ is expected to serve as the integration contour, or equivalently, to define a pOBC-based spectral winding number that can be non-trivial. 
The pOBC spectrum for $N=60$ denoted by red ``$\times$'' in Fig.~\ref{Fig1}(b) approximately divides the complex frequency plane into external and internal domains. Without rigorous proof, one may conjecture that $W_{\text{fGBZ1},\omega_1} = -1$ and $W_{\text{fGBZ1},\omega_2}=0$, where $\omega_c=0.233i$ [black dashed line in Fig.~\ref{Fig1}
(b)] is the critical frequency.  

Further calculation of the dependence of the Green's function on $\omega$ in Fig.~\ref{Fig1}(f) confirms this view.
As the frequency increases, $G_{N,1}(\omega)$ firstly remains independent of frequency, followed by a sharp rise at critical frequency $\omega_c$, and then decreases rapidly beyond that critical frequency, while the dependence of $G_{1,N}(\omega)$ on $\omega$ appears as a mirror curve of that of $G_{N,1}(\omega)$. This also indicates the scale-free amplification and suppression are topological protected.  However, the discrete spectrum alone is generally insufficient for a precise topological characterization, as the fGBZ formalism does not allow for the direct application of equation~\eqref{eq-secII-winding} to the pOBC case. In the next section, We overcome this limitation  by introducing a continuous generalized Brillouin zone (cGBZ) and derive the underlying physics of scale-free response.

\begin{figure}[t]
    \centering
    \includegraphics[width=1\linewidth]{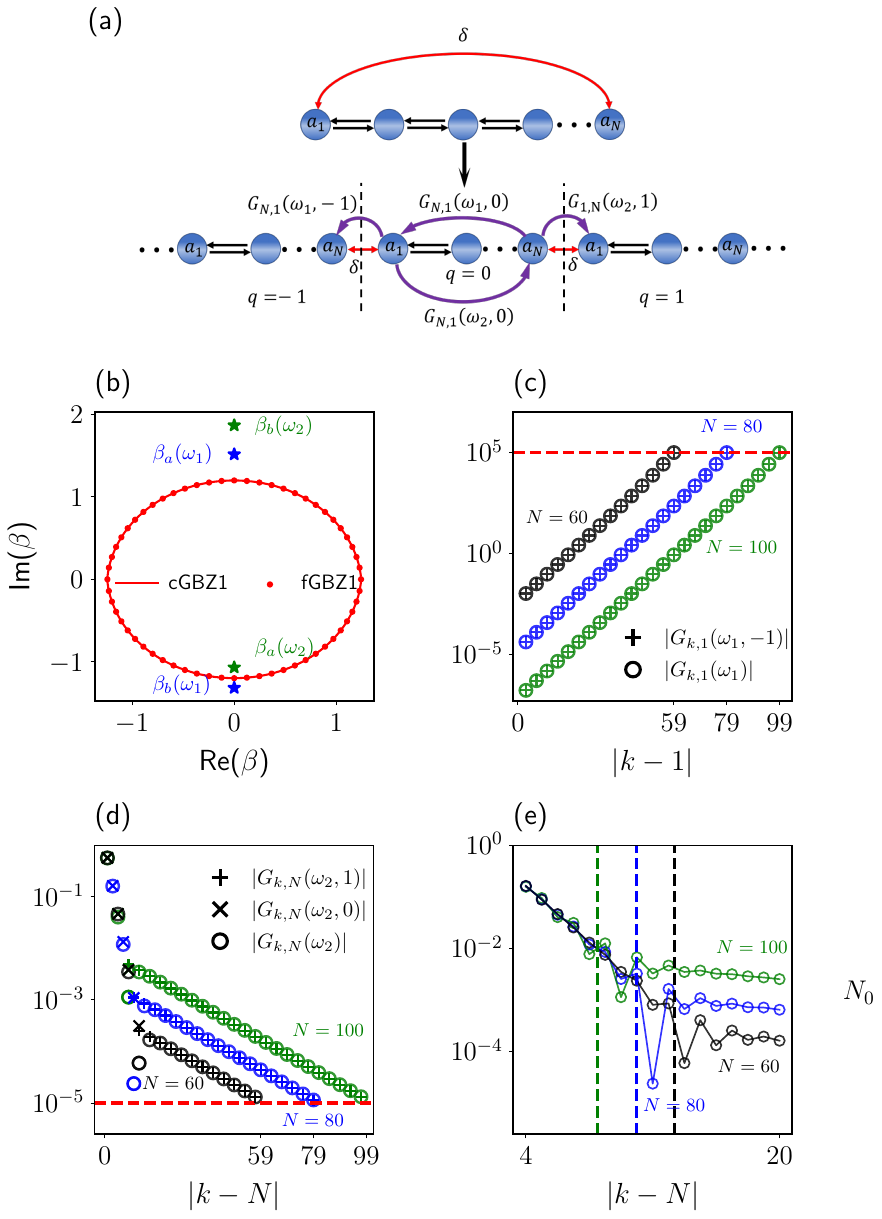}
    \caption{ (a) The pOBC system is equivalent to an infinite chain of $N$-site segments coupled via $\delta$, labeled by $q$ (position relative to the original chain $q=0$).
    (b) Numerical solution of Eq.~\eqref{SecIII-eq-rwave} yields the cGBZ1. The roots of $E(\beta)-\omega=0$ relative to the cGBZ1 are analyzed for $\omega_1$ and $\omega_2$. (c) The numerical result $\abs{G_{k,1}(\omega_1)}$  agrees with the analytic result $\abs{G_{k,1}(\omega_1,-1)}$ and exhibits a scale-free amplification $1/\delta=10^5$ (red dashed line) for $N=60, 80, 100$.  (d) The numerical result $\abs{G_{k,N}(\omega_2)}$  agrees with analytic result  $\abs{G_{k,N}(\omega_2,1)}$ for site $k$ far away from the excitation site and $\abs{G_{k,N}(\omega_2,0)}$ near the excitation site. The $G_{1,N}(\omega_2)$ exhibits a scale-free suppression around $\delta=10^{-5}$ (red dashed line) for $N=60,80,100$. (e) A detailed view of the curve for $|k-N| = 4$ to $20$ in (d) shows the critical distance $N_0$ for different system sizes. The parameters $t_1$, $t_2$, $\delta$, $\omega_1$, and $\omega_2$ are the same as in Fig.~\ref{Fig1}.} 
    \label{Fig2}
\end{figure}

\textit{The continuous generalized Brillouin zone—}For the Hatano-Nelson model with pOBC, the eigenstate with energy $E$ can be written as the superposition of  $\beta_1$ and $\beta_2$ solved by Eq.~\eqref{eq-secII-Hatanobulk}, 
\begin{align}
    \ket{\psi_E} = c_1 \ket{\beta_1}+c_2\ket{\beta_2},\quad 
    \ket{\beta_i} = \frac{1}{\sqrt{N}}\sum_n \beta_i^n \ket{n},
\end{align}
here $\sum_n\ket{n}\bra{n} = I$ and $\bra{n^{\prime}}\ket{n}=\delta_{n^{\prime}n}$. The boundary part of the eigenvalue equation $H\ket{\psi_E}=E\ket{\psi_E}$ gives the constraint,
\begin{align}
\label{eq-secII-boundaryeq}
    \mqty[&-t_2+\delta\beta_1^N &-t_2+\delta\beta_2^N \\
    &\delta\beta_1-t_1\beta_1^{N+1} &\delta\beta_2-t_1\beta_2^{N+1}]\mqty[c_1 \\c_2]=0 .
\end{align}
The non-trivial solution of $c_1,c_2$ requires that the coefficient matrix of equation~\eqref{eq-secII-boundaryeq} has zero determinant，
\begin{align}
\label{eq-SecIII-boundarydet}
    &t_1t_2(\beta_2^{N+1}-\beta_1^{N+1})+\delta^2r^2(\beta_1^{N-1}-\beta_2^{N-1}) \notag \\
    &+(t_2\delta+t_1\delta r^{2N})(\beta_1-\beta_2)=0 ,
\end{align}
with $r^2=t_2/t_1$. In reference~\cite{PhysRevLett.127.116801},  when the system size exceeds the  critical length $N_c=\abs{\log(t_2/\delta)/\log (r)}$, the fGBZ deviates from the GBZ, and the pOBC spectrum becomes complex. The following discussions for the scale-free response are all in the range $N>N_c$. Let $\beta_2 = r x$ and $\beta_1 = r /x$ with $\abs{x}>1$, and consider the leading term in equation~\eqref{eq-SecIII-boundarydet}, we have
\begin{align}
\label{eq-SecIII-iteration1}
    x^{N}=(\delta/t_2)r^N(1-x^{-2}).
\end{align}
From the above equation, the renormalized modes $\tilde{\beta}_1$ and $\tilde{\beta}_2$ can be defined as 
\begin{align}
\label{SecIII-eq-rwave}
    \tilde{\beta}_1=\beta_1 \sqrt[N]{(1-\beta_1^2/r^2)} , \quad
    \tilde{\beta}_2=\frac{\beta_2}{\sqrt[N]{1-r^2/\beta_2^2}}.
\end{align}
Consequently, we introduce two maps, $f_1:\beta_1 \rightarrow \tilde{\beta}_1 $ and $f_2:\beta_2 \rightarrow \tilde{\beta}_2 $ which are both one to one mappings. The renormalized modes $\tilde{\beta}_1=\sqrt[N]{(t_2/\delta)}e^{i\theta}$ and $\tilde{\beta}_2=r^2\sqrt[N]{(\delta/t_2)}e^{-i\theta}$ with $\theta=2m\pi/N$ and $m\in \mathcal{Z}$ are always confined to two circles on the complex plane, respectively. The variable $\theta$ plays a role analogous to the lattice momentum, and taking the continuum limit in $\theta$ defines the renormalized generalized Brillouin zone (rGBZ), which consists of two branches labeled rGBZ1 and rGBZ2. Employing the inverse maps $f_1^{-1}$ and $f_2^{-1}$, we extend their preimages $\tilde{\beta}_1$ and $\tilde{\beta}_2$ across the entire rGBZ1 and rGBZ2 curves yielding two smooth curves that invariably encompass fGBZ1 and fGBZ2. We interpret these curves as a continuous extension of the fGBZ, referring to them as the continuous generalized Brillouin zone (cGBZ). Correspondingly, these two curves are denoted as cGBZ1 and cGBZ2. In Fig.~\ref{Fig2}(b), we numerically solve Eq.~\eqref{SecIII-eq-rwave} by replacing $\tilde{\beta}_1$ with points on the rGBZ1 , which yields a smooth curves cGBZ1.

As for the bulk modes, $\beta_1$ in cGBZ1 dominants the localized behavior of the system. The localization length can generally be calculated as~\cite{PhysRevB.104.165117,li2021impurity,li2020critical}
\begin{align}
    L_c=\abs{1/\log(\abs{\beta_1})}\simeq N\abs{\log (\delta/t_2)}.
\end{align}
The localization length $L_c$ is proportional to the size of system, a phenomenon known as scale-free localization~\cite{li2021impurity,li2020critical}, which is different from the conventional skin modes. 
When scale-free localization exists, the pOBC spectrum differs from the OBC spectrum and can exhibit a nontrivial point-gap topology~\cite{doi:10.1080/00018732.2021.1876991,PhysRevX.8.031079}, characterized by a non-vanishing winding number $W_{\text{cGBZ1},\omega}$ defined on cGBZ1. The dynamical properties of the system are closely related to the localization behavior of its eigenstates~\cite{zirnstein2021exponentially,zirnstein2021bulk,PhysRevLett.124.056802}. Analogous to the skin effect, the emergence of scale-free amplification can be attributed to the scale-free localization in the system. Meanwhile, the winding number $W_{\text{cGBZ1},\omega}$ is also anticipated to indicate whether the  response at frequency $\omega$ is of the scale‑free amplification type. In the next section, we will give an analytic discussion.

\textit{The analytic calculation of scale-free response—}In the biorthogonal basis, the Green's function takes the form,
\begin{align}
\label{eq-SECIII-generalgf}
    G(\omega)=\sum_{E} \frac{\ket{\psi_{E}}\bra{\tilde{\psi}_E}}{\bra{\tilde{\psi}_E}\ket{\psi_{E}}} \frac{1}{\omega-E}.
\end{align}
The left eigenvector of $\ket{\psi_E}$ is defined by $H^{\dagger}\ket{\tilde{\psi}_E}=E^{\dagger}\ket{\tilde{\psi}_E}$~\cite{doi:10.1080/00018732.2021.1876991}, with $\ket{\tilde{\psi}_E} = \tilde{c}_1 \ket{(\beta_1^{\ast})^{-1}}+\tilde{c}_2\ket{(\beta_2^{\ast})^{-1}}$. This leads to a boundary equation similar to the equation~\eqref{eq-secII-boundaryeq},
\begin{align}
\label{eq-secIII-boundaryeq}
    \mqty[&-t_1^{\ast}+\delta^{\ast}(\beta_1^{\ast})^{-N} &-t_1^{\ast}+\delta^{\ast}(\beta_2^{\ast})^{-N} \\
    &\delta^{\ast}\beta_1^{\ast}-t_2^{\ast}(\beta_1^{\ast})^{-N-1} &\delta^{\ast}\beta_2^{\ast}-t_2^{\ast}(\beta_2^{\ast})^{-N-1}]\mqty[\tilde{c}_1 \\ \tilde{c}_2]=0 .
\end{align}
From the boundary equation~\eqref{eq-secII-boundaryeq} and ~\eqref{eq-secIII-boundaryeq}, the coefficients of eigenvectors are solved as $\tilde{c}_1 \simeq -\tilde{c}_2$ and
$c_1/c_2 \simeq -\beta_2^{N+1}/\beta_1^{N+1}$. Assuming that $c_1=\tilde{c}_1=1$, the Green's function can be written as
\begin{align}
\label{eq-SECIII-sum1gf}
    G_{k,l}(\omega)=\!\!\!\!\!\!\sum_{\beta_1 \in \text{fGBZ1}}\! \! \! \! \frac{\beta_1^{k-l}-\frac{\beta_1^{k+l}}{r^{2l}}-(\frac{\beta_1^{2N+2-k-l}}{r^{2N+2-2k}}-\frac{\beta_1^{2N+2+l-k}}{r^{2N+2+2l-2k}})}{(N-\frac{2\beta_1^2/r^2}{1-\beta_1^2/r^2})(\omega-E(\beta_1))} .\end{align}
The numerator of  equation~\eqref{eq-SECIII-sum1gf} consists of four terms. The first and last terms come from the contribution of the states $\ket{\beta_1}$ and $\ket{\beta_2}$ individually.  By replacing the summation with an integral along the GBZ, the first and last terms converge to the Green's function given in previous works~\cite{Wangzhong2021Simpleformulas}. The second and third terms arise from the boundary effect, resulting in the coupling of the states $\ket{\beta_1}$ and $\ket{\beta_2}$~\cite{PhysRevB.109.205407}, which are generally negligible~\cite{PhysRevB.109.205407,Wangzhong2021Simpleformulas,PhysRevB.107.115412}. However, in pOBC case, due to the sensitivity of the boundary condition, these terms become important, thereby contributing to the Green's function. Meanwhile, the size dependence of the pOBC spectrum also makes the approximation of replacing the summation with an integral in the thermodynamic limit occasionally unreliable. Such concerns imply that the Green's function under pOBC can't be obtained simply by changing the integration loop to the cGBZ, as done for OBC.

To address the above issues, we expand equation~\eqref{eq-SECIII-sum1gf} as a Laurent series with respect to $\tilde{\beta}_1$ on the rGBZ1 and replace $\tilde{\beta}_1$ with $\beta_1$ on the cGBZ1.
\begin{align}
\label{eq-SECIII-generalizedexpression}
&G_{k,l}(\omega)=\!  \!\sum_{q=-\infty}^{\infty} \! G_{k,l}(\omega,q) = \! \! \sum_{q=-\infty}^{\infty} (\frac{\delta}{t_2})^q \oint_{\beta_1 \in \text{cGBZ1} } \notag \\  &\frac{\beta_1^{k-l}-\frac{\beta_1^{k+l}}{r^{2l}}-(\frac{\beta_1^{2N+2-k-l}}{r^{2N+2-2k}}-\frac{\beta_1^{2N+2+l-k}}{r^{2N+2+2l-2k}})}{2\pi i\beta_1(\omega-E(\beta_1))\beta_1^{-qN}(1-\beta_1^2/r^2)^{-q}}  d\beta_1.
\end{align} 
The detailed calculation of the expansion can be found in  Supplemental Material, Sec. I. Here, we provide an intuitive explanation for Eq.~\eqref{eq-SECIII-generalizedexpression}. As shown in Fig.~\ref{Fig2}(a), this model can be viewed as a chain composed of infinitely many identical $N$-site subchains, labeled by an integer $q$ (where $q = 0$ corresponds to the chain with an initial excitation), coupled to each other by a hopping amplitude $\delta$. $G_{k,l}(\omega,q)$ describes the response at site $k$ in the $q$-chain to an excitation at site $l$ in the $0$-chain. Here, positive $q$ corresponds to a hopping from $a_{n}$ to $a_{1}$ with $q$ times, which can be regarded as a $q$‑order right boundary effect; negative $q$ corresponds to the opposite direction (i.e., from $a_1$ to $a_n$). For the model illustrated in Fig.~\ref{Fig2}(a), the total response at the physical site $k$ is obtained by summing contributions from the corresponding site $k$ across all $q$-chains. In different topological regimes of $W_{\text{cGBZ1},\omega}$, $G_{k,l}(\omega)$ is dominated by terms $G_{k,l}(\omega,q)$ with different order boundary effects, leading to different types of response in the system. 

For the frequency $\omega=\omega_1$  in the internal domain of the pOBC spectrum, $W_{\text{cGBZ1},\omega_1}$ is non-trivial. In this regime, both roots of $E(\beta)-\omega_1=0$ (ordered by $|\beta_a| \le |\beta_b|$) lie outside cGBZ1 [blue asterisks in Fig.~\ref{Fig2}(b)]. The integral in equation~\eqref{eq-SECIII-generalizedexpression} is solved in Supplemental Material, Sec. II. When $k \geq l$,  we have $G_{k,l}(\omega_1) \simeq G_{k,l}(\omega_1,-1) $, as verified in Fig.~\ref{Fig2}(c), with analytic result $|G_{k,1}(\omega_1,-1)|$ matching  numerical result $|G_{k,1}(\omega_1)|$. The end-to-end Green's function can be obtained by taking the residue theorem,
\begin{align}\label{eq-endto}
    G_{N,1}(\omega_1)=\frac{t_2}{\delta}(\frac{\beta_a^{-1}}{ t_1(\beta_a-\beta_b)}   +  \frac{\beta_b^{-1}}{ t_1(\beta_b-\beta_a)}  )=-\frac{1}{\delta},
\end{align}
which is only determined by the perturbation. As shown in Fig.~\ref{Fig2}(c), for different system sizes, although $G_{k,1}(\omega_1)$ increases exponentially with the distance from the excitation point, the end to end $ \abs{G_{N,1}(\omega_1)}$ exhibits a uniform value of $1/\delta=10^5$. This independence of the end‑to‑end Green's function in Eq.~\eqref{eq-endto} from system size, frequency, and even coupling strength  renders simplifying device design and easy to control the signal amplification in realistic implementations. 
In the opposite direction  $k<l$, the Green's function reads $G_{k,l}(\omega_1) \simeq G_{k,l}(\omega_1,0) \propto \beta_a^{-N}$, exhibiting size‑dependent exponential suppression.

For the frequency $\omega=\omega_2$ in the external domain of pOBC spectrum,  the $W_{\text{cGBZ1},\omega_2}$ is trivial.  In this regime, $\beta_a$ lies inside cGBZ1, while $\beta_b$ lies outside cGBZ1 [green asterisks in Fig.~\ref{Fig2}(b)]. The corresponding integral in equation~\eqref{eq-SECIII-generalizedexpression} is solved in Supplemental Material, Sec. III. As seen in Fig.~\ref{Fig2}(d), when $k<l$, such as $l=N$, the numerical result $\abs{G_{k,N}(\omega_2)}$ agrees with the analytic result  $\abs{G_{k,N}(\omega_2,1)}$ at large distances $\abs{k-N}$ and with $\abs{G_{k,N}(\omega_2,0)}$ at small distance $\abs{k-N}$. The critical length is labeled as $N_0$. Here the critical distance $N_0$ is estimated as $\abs{(N\log \abs{\beta_a}+\log \delta)/(\log \abs{\beta_a}-\log\abs{\beta_b)}}$. This anomalous scale behavior is also discussed in reference~\cite{d5zc-p1sk} and Supplemental Material, Sec.~IV. In Fig.~\ref{Fig2}(e), the theoretical critical distance $N_0$ for different system sizes are indicated by dashed lines, which agree well with the transition positions on the numerical curves. For the case $k=1$ and $l=N$, the end to end Green's function
\begin{align}
    G_{1,N}(\omega_2)&= -\frac{\delta}{t_2} \frac{\beta_a-\frac{\beta_a^3}{2}}{t_1(\beta_a-\beta_b)} \simeq \frac{-\delta}{t_2t_1},
\end{align}
which exhibits scale-free suppression independent of system size, as shown in Fig.～\ref{Fig2}(d). In the opposite direction $k \geq l$,  we have  $G_{k,l}(\omega_1) \simeq G_{k,l}(\omega_1,0) \propto \beta_a^{N}$, exhibiting size-dependent exponential amplification when $\omega_2$ lies in the PBC spectrum. From $G_{N,1}(\omega_1) \simeq G_{N,1}(\omega_1,-1)$ and $G_{1,N}(\omega_2)=G_{1,N}(\omega_2,1)$, we can conclude that the scale-free response is  caused by the first order  boundary effect and captured by the $W_{\text{cGBZ1},\omega}$.

\begin{figure}
    \centering
    \includegraphics[width=1\linewidth]{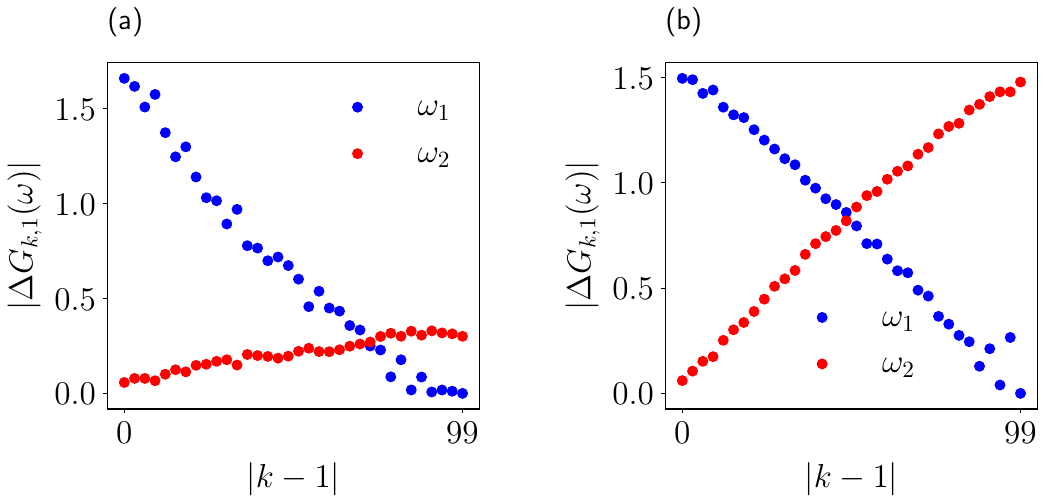}
    \caption{Relative error of the Green's function under uniform distribution of disorder $\epsilon \sim U(-0.05,0.05)$ on (a) hoppings  and (b) on-site terms. Other parameters are as in Fig.~\ref{Fig1}, except $\omega_1 = 0.25i$.
    \label{fig3}}
\end{figure}

Lastly, we address the  robustness of this scale‑free response against disorder in the system parameters. We perturb the hopping terms and on-site energy with a uniform distribution of disorder $\epsilon \sim U(-0.05,0.05)$. To quantify the influence of the disorder, we define the relative error of the Green's function as
\begin{align}
    \Delta G_{k,l}(\omega) = \frac{\bar{G}_{k,l}(\omega)-G_{k,l}(\omega)}{G_{k,l}(\omega)},
\end{align}
where $\bar{G}_{k,l}(\omega)$ is the Green's function with disorder. Since the amplification of response grows exponentially with distance, the end to end  response is of particular interest~\cite{Wangzhong2021Simpleformulas,PhysRevB.105.045122}. When $\omega = \omega_1$, $\Delta G_{k,1}(\omega)$ tends to zero as the observation site moves away from the initial site, rendering the end‑to‑end scale‑free amplification highly robust, as shown in Fig.~\ref{fig3}(a) and (b), regardless of whether the disorder is added to the onsite or hopping terms. In contrast, when $\omega = \omega_2$, the amplification becomes size‑dependent, which induces a maximum error in the end‑to‑end Green's function in the presence of perturbation. The robustness inherent in the scale‑free response makes it particularly promising for experimental realization, as it requires no fine‑tuning of system parameters.

\textit{Conclusion–}In this work, we propose a new type response — scale‑free amplification or suppression. The scale‑free amplification can be simultaneously topological, scale‑free, and directionally amplified, rendering it potentially applicable. By making the fGBZ1 continuous into the cGBZ1, we define a winding number $W_{\text{cGBZ1},\omega}$ that characterizes the scale-free response. An intuitive explanation for the scale-free response can be given by considering the  first order boundary effect. Such scale-free amplification is robust against disorder in the system parameters, which makes observation and application of our proposed response easier. Recently, several new studies on non-Bloch response have emerged,   providing a way to measure the Green’s function at complex frequencies~\cite{2024arXiv241112577H,2026CmPhy...9...84H}. These developments also make our finding experimentally feasible. 

\vspace{10pt}

 ~\textit{Acknowledgments—}This work is supported by the
Natural Science Foundation of Hunan Province (Grant No. 2024JJ6011) and Quantum Science and Technology-National Science and Technology Major Project (Grant No. 2021ZD0302300).

\bibliography{manuscript.bib}

\end{document}


\title{Supplemental Materials for ``Scale-Free Response with Directional Amplification in Critical Non-Hermitian Systems"}

\author{Kunling Zhou}
\affiliation{School of Physics, Huazhong University of Science and Technology, Wuhan 430074, P. R. China}

\author{Zihe Yang}
\affiliation{School of Physics, Huazhong University of Science and Technology, Wuhan 430074, P. R. China}

\author{Bowen Zeng}
\email[]{zengbowen@csust.edu.cn}
\affiliation{Hunan Provincial Key Laboratory of Flexible Electronic Materials Genome Engineering,
School of Physics and Electronic Sciences, Changsha University of Science and Technology, Changsha 410114, P. R. China}

\author{Yong Hu}
\email[]{huyong@hust.edu.cn}
\affiliation{School of Physics, Huazhong University of Science and Technology, Wuhan 430074, P. R. China}

\makeatletter
\newcommand{\rmnum}[1]{\romannumeral #1}
\newcommand{\Rmnum}[1]{\expandafter\@slowromancap\romannumeral #1@}
\makeatother

\maketitle

\section{The expansion of the Green's function}

In this section, we will give a detailed calculation for the Laurent expansion of the Green's function in the main text. The renormalized wavevector $\tilde{\beta}_1$ is defined by taking the map,
\begin{align}
\label{eq-sup1-newwavevector1}
    f_1:\tilde{\beta}_1=\beta_1 \sqrt[N]{(1-\beta_1^2/r^2)} .
\end{align}
Here, $\tilde{\beta}_1 = \sqrt[N]{t_2/\delta} \, e^{m 2\pi i / N}$, and the set of $\tilde{\beta}_1$ lies on a circle of radius $\sqrt[N]{t_2/\delta}$, which is referred to as the rGBZ1. Expand the preimage of $f_1^{-1}$ to the whole rGBZ1,  $f_1^{-1}$ maps arbitrary point $\tilde{\beta}_1'$ on rGBZ1 to $\beta_1'$ by the relation,
\begin{align}\label{eq-sup1-newwavevector}
\tilde{\beta}_1^{\prime}=\beta_1^{\prime} \sqrt[N]{(1-(\beta_1^{\prime})^2/r^2)}.
\end{align}
The set of $\beta_1'$ is a smooth curve contains the fGBZ1 and denoted as cGBZ1. The differential of the equation~\eqref{eq-sup1-newwavevector} is
\begin{align}
\label{eq-sup1-differential1}
    d \tilde{\beta}_1^{\prime} = d\beta_1^{\prime} \sqrt[N]{(1-(\beta_1^{\prime})^2/r^2)}(1-\frac{2(\beta_1^{\prime})^2/r^2}{N(1-(\beta_1^{\prime})^2/r^2)}) .
\end{align}
We expand the term 
\begin{align}
    G_{k,l}(\omega_1)=\sum_{\beta_1}\frac{\beta_1^{k-l}-\frac{\beta_1^{k+l}}{r^{2l}}-(\frac{\beta_1^{2N+2-k-l}}{r^{2N+2-2k}}-\frac{\beta_1^{2N+2+l-k}}{r^{2N+2+2l-2k}})}{(N-\frac{2\beta_1^2/r^2}{1-\beta_1^2/r^2})(\omega-E(\beta_1))}
\end{align}
in a Laurent expansion $\sum_{\tilde{\beta}_1}\sum_{n=-\infty}^{\infty} a_n\tilde{\beta}_1^n$ with respect to the $\tilde{\beta}_1^{\prime}$ on the rGBZ1~\cite{Wangzhong2021Simpleformulas}, 
\begin{align}
\label{eq-sup1-expandcoeffi}
    a_n=  \oint_{\tilde{\beta}_1^{\prime}\in \text{rGBZ1}} \frac{\beta_1^{k-l}-\frac{\beta_1^{k+l}}{r^{2l}}-(\frac{\beta_1^{2N+2-k-l}}{r^{2N+2-2k}}-\frac{\beta_1^{2N+2+l-k}}{r^{2N+2+2l-2k}})}{2\pi i(N-\frac{2\beta_1^2/r^2}{1-\beta_1^2/r^2})(\omega-E(\beta_1))}  (\tilde{\beta}_1^{\prime})^{-n-1} d\tilde{\beta}_1^{\prime}.
\end{align}
For convenience, we use $\beta_1$ to replace the $\beta_1^{\prime}$ in the equation~\eqref{eq-sup1-expandcoeffi}, and this notation is used in the main text and the following sections. The Green's function can be rewritten as
\begin{align}
G_{k,l}(\omega)= \sum_{\tilde{\beta}_1}\sum_{n=-\infty}^{\infty}\tilde{\beta}_1^n\oint_{\tilde{\beta}_1^{\prime}\in \text{rGBZ1}} \frac{\beta_1^{k-l}-\frac{\beta_1^{k+l}}{r^{2l}}-(\frac{\beta_1^{2N+2-k-l}}{r^{2N+2-2k}}-\frac{\beta_1^{2N+2+l-k}}{r^{2N+2+2l-2k}})}{2\pi i(N-\frac{2\beta_1^2/r^2}{1-\beta_1^2/r^2})(\omega-E(\beta_1))}  (\tilde{\beta}_1^{\prime})^{-n-1} d\tilde{\beta}_1^{\prime}. 
\end{align}
The orthogonality relation for $\tilde{\beta}_1$ is $\sum_{\tilde{\beta}_1}\tilde{\beta}_1^n=\sum_{m} (\sqrt[N]{(t_2/\delta)})^ne^{2mn\pi i /N}=N(t_2/\delta)^q \delta_{n,qN} $ and $q \in \mathcal{Z}$. According to the orthogonality and taking $q \rightarrow -q$, the Green's function becomes
\begin{align}
G_{k,l}(\omega)= \! \! \sum_{q=-\infty}^{\infty}(\frac{\delta}{t_2})^q\oint_{\tilde{\beta}_1^{\prime}\in \text{rGBZ1}} \frac{\beta_1^{k-l}-\frac{\beta_1^{k+l}}{r^{2l}}-(\frac{\beta_1^{2N+2-k-l}}{r^{2N+2-2k}}-\frac{\beta_1^{2N+2+l-k}}{r^{2N+2+2l-2k}})}{2\pi i(N-\frac{2\beta_1^2/r^2}{1-\beta_1^2/r^2})(\omega-E(\beta_1))}  (\tilde{\beta}_1^{\prime})^{qn-1} d\tilde{\beta}_1^{\prime} .
\end{align}
Using the  equation~\eqref{eq-sup1-differential1} and equation~\eqref{eq-sup1-newwavevector1}, We replace the $\tilde{\beta}_1^{\prime}$ in the rGBZ1 with the $\beta_1$ in the cGBZ1, 
\begin{align}
\label{eq-sup1-expanGreen}
G_{k,l}(\omega)= \! \! \sum_{q=-\infty}^{\infty} \! G_{k,l}(\omega,q) = \sum_{q=-\infty}^{\infty} (\frac{\delta}{t_2})^q \oint_{\beta_1 \in \text{cGBZ1}} \frac{\beta_1^{k-l}-\frac{\beta_1^{k+l}}{r^{2l}}-(\frac{\beta_1^{2N+2-k-l}}{r^{2N+2-2k}}-\frac{\beta_1^{2N+2+l-k}}{r^{2N+2+2l-2k}})}{\beta_1(\omega-E(\beta_1))} \frac{\beta_1^{qN}(1-\beta_1^2/r^2)^{q} d\beta_1}{2\pi i}.
\end{align}
This result shows that the Green's function for the system with pOBC can be written as a summation of integrals which contains term $q=0$ without boundary effect, $q>0$ with right boundary effect and $q<0$ with left boundary effect.

\section{topological non-trivial regime}

In this section, we discuss the leading term in the summation of the Green's function $G_{k,l}(\omega)$ for the case where $\omega = \omega_1$ lies in the topologically nontrivial region with $W_{\text{cGBZ1},\omega_1} = -1$. The two roots of $E(\beta) - \omega=0$, ordered by $|\beta_a| \leq |\beta_b|$, lie outside the cGBZ1. The discussion can be divided into three parts, corresponding to the terms without boundary effect, with right boundary effect, and with left boundary effect.

Firstly, we consider the $q = 0$ part without boundary effect
\begin{align}
G_{k,l}(\omega_1,0) =  \oint_{\beta_1 \in \text{cGBZ1}} \frac{\beta_1^{k-l}-\frac{\beta_1^{k+l}}{r^{2l}}-(\frac{\beta_1^{2N+2-k-l}}{r^{2N+2-2k}}-\frac{\beta_1^{2N+2+l-k}}{r^{2N+2+2l-2k}})}{-2\pi i t_1(\beta_1-\beta_a)(\beta_1-\beta_b)}  d\beta_1 .
\end{align}
When $k \geq l$, the pole $\beta_a$, $\beta_b$, and $\infty$ are all out of the cGBZ1, which leads to 
\begin{align}
    G_{k,l}(\omega_1,0)=0.
\end{align}
When $k<l$, the pole $0$ is now in the cGBZ1 and thus,
\begin{align}
    G_{k,l}(\omega_1,0) = \frac{\beta_a^{k-l}}{ t_1(\beta_a-\beta_b)}   +  \frac{\beta_b^{k-l}}{ t_1(\beta_b-\beta_a)} .
\end{align}

The right boundary effect for $q>0$ can be considered in the same way. For arbitrary $0<k,l<N$, there is always no pole in the cGBZ1 and this part takes no contribution, 
\begin{align}
&\sum_{q=1}^{\infty} G_{k,l}(\omega_1,q) =\sum_{q=1}^{\infty} (\frac{\delta}{t_2})^q \oint_{\beta_1 \in \text{cGBZ1}} \frac{\beta_1^{k-l}-\frac{\beta_1^{k+l}}{r^{2l}}-(\frac{\beta_1^{2N+2-k-l}}{r^{2N+2-2k}}-\frac{\beta_1^{2N+2+l-k}}{r^{2N+2+2l-2k}})}{-2\pi i t_1 (\beta_1-\beta_a)(\beta_1-\beta_b)} \beta_1^{qN}(1-\beta_1^2/r^2)^{q} d\beta_1 =0  . 
\end{align}

The left boundary effect for $q<0$ is different, since the points $\pm r$ can be poles of the system. To determine the order of the poles  $\pm r$, we take a transformation,
\begin{align}
\label{eq-sup2-transformationG}
    &1-\frac{\beta_1^{2l}}{r^{2l}}=(1-
    \frac{\beta_1^{2}}{r^2})+(
    \frac{\beta_1^{2}}{r^2}-\frac{\beta_1^{4}}{r^4}) 
    +\cdots+(
    \frac{\beta_1^{2l-2}}{r^{2l-2}}-\frac{\beta_1^{2l}}{r^{2l}})=\sum_{m=0}^{l-1}\frac{\beta_1^{2m}}{r^{2m}}(1-\frac{\beta_1^2}{r^2})    .
\end{align}
This part thus becomes that 
\begin{align}
\label{eq-int2 gf}
      &G_{k,l}(\omega_1)= \! \!\sum_{q=-\infty}^{-1} \! G_{k,l}(\omega_1,q) = \! \! \sum_{q=-\infty}^{-1} (\frac{\delta}{t_2})^q\oint_{\beta_1 \in \text{cGBZ1}}  \frac{\sum_{m=0}^{l-1}(\frac{\beta_1^{2m}}{r^{2m}}-\frac{\beta_1^{2N+2-2k+2m}}{r^{2N+2-2k+2m}})\beta_1^{k-l}}{-2\pi i t_1(\beta_1-\beta_a)(\beta_1-\beta_b)}   \beta_1^{qN}(1-\beta_1^2/r^2)^{q+1}d\beta_1 \notag \\
      &=\! \! \sum_{q=-\infty}^{-1} (\frac{\delta}{t_2})^q\oint_{\beta_1 \in \text{cGBZ1}}  \frac{\sum_{m=0}^{l-1}\sum_{m_1=0}^{N-k}\frac{\beta_1^{2m+2m_1+k-l}}{r^{2m+2m_1}}}{-2\pi i t_1(\beta_1-\beta_a)(\beta_1-\beta_b)} \beta_1^{qN}(1-\beta_1^2/r^2)^{q+2}d\beta_1 .
\end{align}
Consequently,  the $\pm r$ are not poles of $G_{k,l}(\omega_1,q)$ for $q=-1,-2$. In the topological non-trivial regime, the roots $\beta_a$ and $\beta_b$ satisfy
\begin{align}
   \abs{\frac{1}{\beta_{b}^{N}\delta}}\ll   \abs{\frac{1}{r^{N}\delta}} \ll   \abs{\frac{1}{\beta_{a}^{N}\delta}} \ll 1 .
\end{align} Taking the residue theorem, we can find that for $q<0$, the term $G_{k,l}(\omega_1,q)$ converge as $(1/\delta)^q \beta_a^{k-l-qN}$ and the term $G_{k,l}(\omega_1,-1)$  will be the leading term for the part $q>0$. Assume that $l_1=\min\{l-1,[(N+l-1-k)/2]\}$, $l_2=\min\{l-1,[(k+l-N-3)/2]\}$ , and the $ G_{k,l}(\omega_1,-1)$ becomes,
\begin{align}
   G_{k,l}(\omega_1,-1)= \frac{t_2}{\delta}\frac{\sum_{m=0}^{l_1}(\frac{\beta_a^{k-l+2m-N}}{r^{2m}}-\frac{\beta_b^{k-l+2m-N}}{r^{2m}})}{t_1(\beta_a-\beta_b)} -\frac{t_2}{\delta}\frac{\theta(l_2)\sum_{m=0}^{l_2}(\frac{\beta_a^{N+2-k+2m-l}}{r^{2N+2-2k+2m}}-\frac{\beta_b^{N+2-k+2m-l}}{r^{2N+2-2k+2m}})}{t_1(\beta_a-\beta_b)} .
\end{align}
Here, $\theta(l_2)$ is the step function.

Now we can conclude that for $k \geq l$, since  $\sum_{q=1}^{\infty}G_{k,l}(\omega_1,q)=0$ and $G_{k,l}(\omega_1,0)=0$,  $G_{k,l}(\omega_1,-1)$ will be the dominant term. While for the case $k<l$ , the term $q=0$ contributes to the $G_{k,l}(\omega)$,
and it will be the leading term replace to the term $G_{k,l}(\omega_1,-1)$. The Green's function can be written as,
\begin{align}
\label{eq-sup-non-trivialgreenf}
    G_{k,l}(\omega_1)\simeq\left\{\begin{array}{l}
G_{k,l}(\omega_1,-1)=\frac{t_2}{\delta}\oint_{\beta_1 \in \text{cGBZ1}}  \frac{\sum_{m=0}^{l-1}(\frac{\beta_1^{k-l+2m-N}}{r^{2m}}-\frac{\beta_1^{N+2-k+2m-l}}{r^{2N+2-2k+2m}})}{-2\pi i t_1(\beta_1-\beta_a)(\beta_1-\beta_b)} d\beta_1, k \geqslant l, \\
G_{k,l}(\omega_1,0) =  \oint_{\beta_1 \in \text{cGBZ1}} \frac{\beta_1^{k-l}-\frac{\beta_1^{k+l}}{r^{2l}}-(\frac{\beta_1^{2N+2-k-l}}{r^{2N+2-2k}}-\frac{\beta_1^{2N+2+l-k}}{r^{2N+2+2l-2k}})}{-2\pi i t_1(\beta_1-\beta_a)(\beta_1-\beta_b)}  d\beta_1 , k < l .
\end{array}\right.
\end{align}
Especially, $G_{N,1}(\omega_1) \simeq -1/\delta$ which directly leads to the scale-free amplification of the system. For the opposite direction response, $G_{1,N}(\omega_1) \simeq (\beta_a^{1-N}-\beta_b^{1-N})/(\beta_a-\beta_b)$ which will lead to a size-dependent suppression scaling $\beta_a^{-N}$.

\section{topological trivial regime}
We now turn to the case where $\omega = \omega_2$ is in the topologically trivial region, $W_{\text{cGBZ1},\omega_2} = 0$. Here, the root $\beta_a$ lies inside the cGBZ1, whereas $\beta_b$ lies outside.

Considering the part without boundary effect for $q=0$, the integral can be calculated in two cases. The first case is when $k < l$,
\begin{align}
&G_{k,l}(\omega_2,0)=  \oint_{\beta_b \in \text{cGBZ1}} \frac{\beta_1^{k-l}-\frac{\beta_1^{k+l}}{r^{2l}}-(\frac{\beta_1^{2N+2-k-l}}{r^{2N+2-2k}}-\frac{\beta_1^{2N+2+l-k}}{r^{2N+2+2l-2k}})}{- 2\pi i t_1  (\beta_1-\beta_a)(\beta_1-\beta_b)}  d\beta_1  \notag \\
&= \frac{\beta_b^{k-l}}{t_1(\beta_b-\beta_a)}+\frac{\frac{\beta_a^{k+l}}{r^{2l}}+(\frac{\beta_a^{2N+2-k-l}}{r^{2N+2-2k}}-\frac{\beta_a^{2N+2+l-k}}{r^{2N+2+2l-2k}})}{t_1(\beta_a-\beta_b)} \simeq  \frac{\beta_b^{k-l}}{t_1(\beta_b-\beta_a)} .
\end{align}
When $k \geq l$, it becomes
\begin{align}
&G_{k,l}(\omega_2,0)=  \oint_{\beta_b \in \text{cGBZ1}} \frac{\beta_1^{k-l}-\frac{\beta_1^{k+l}}{r^{2l}}-(\frac{\beta_1^{2N+2-k-l}}{r^{2N+2-2k}}-\frac{\beta_1^{2N+2+l-k}}{r^{2N+2+2l-2k}})}{- 2\pi i t_1  (\beta_1-\beta_a)(\beta_1-\beta_b)}  d\beta_1  \notag \\
&= -\frac{\beta_a^{k-l}-\frac{\beta_a^{k+l}}{r^{2l}}+(\frac{\beta_a^{2N+2-k-l}}{r^{2N+2-2k}}-\frac{\beta_a^{2N+2+l-k}}{r^{2N+2+2l-2k}})}{t_1(\beta_a-\beta_b)} \simeq  \frac{\beta_a^{k-l}-\frac{\beta_a^{k+l}}{r^{2l}}}{t_1(\beta_b-\beta_a)}  .  
\end{align}

For the left boundary effect part with $q<0$, when $q<-2$ (see Eq.~\ref{eq-int2 gf}), the poles $\beta_b$ and $\pm r$ lie outside the cGBZ1.  In the topological trivial regime,
\begin{align}
   \abs{\frac{1}{\beta_{b}^{N}\delta}}\ll   \abs{\frac{1}{r^{N}\delta}} \ll 1 \ll   \abs{\frac{1}{\beta_{a}^{N}\delta}} .
\end{align}
We get that for  $q<-2$ the term $G_{k,l}(\omega_2,q)$ will converge as $(1/\delta)^q r^{-qN}$ and we ignore these high order terms.  In the following, we will calculate the contribution of $G_{k,l}(\omega_2,-1)$ and $G_{k,l}(\omega_2,-2)$. According to the equation~\eqref{eq-sup2-transformationG}, $G_{k,l}(\omega_2,-1)$ can be written as,
\begin{align}
&G_{k,l}(\omega_2) \simeq G_{k,l}(\omega_2,-1)=\frac{t_2}{\delta}\oint_{\beta_1 \in \text{cGBZ1}}  \frac{(\sum_{m_0=0}^{l-1}\frac{\beta_1^{k-l+2m_0-N}}{r^{2m_0}}-\sum_{m_1=0}^{l-1}\frac{\beta_1^{N+2-k+2m_1-l})}{r^{2N+2-2k+2m_1}}}{-2\pi i t_1(\beta_1-\beta_a)(\beta_1-\beta_b})   d\beta_1,
\end{align}
in which the integral of $G_{k,l}(\omega_2,-1)$  is consisted of two summations. We consider an arbitrary term in  the first summation of $G_{k,l}(\omega_2,-1)$ . If $k-l+2m_0-N \leq 0$, it can be calculated as 
\begin{align}
    \frac{t_2}{\delta}\frac{\frac{\beta_a^{k-l+2m_0-N}}{r^{2m_0}}}{t_1(\beta_b-\beta_a)}=\frac{t_2}{\delta}\frac{\frac{\beta_b^{l-k-2m_0+N}}{r^{2N+2l-2k-2m_0}}}{t_1(\beta_b-\beta_a)}.
\end{align}
Such a term can be canceled with the term $m_1=l-1-m_0$ in the second summation. In the same way, the case $k-l+2m_0-N > 0$ in the first summation will also cancel with the term $m_1=l-1-m_0$ in the second summation. Thus $G_{k,l}(\omega_2,-1)$ takes no contribution to the $G_{k,l}(\omega_2)$. When discussing the contribution of $G_{k,l}(\omega_2,-2)$, we write it as
\begin{align}
&G_{k,l}(\omega_2,-2)= \frac{t_2^2}{\delta^2} \oint_{\beta_1 \in \text{cGBZ1}} \frac{\beta_1^{k-l-2N}-\frac{\beta_1^{k+l-2N}}{r^{2l}}-(\frac{\beta_1^{2-k-l}}{r^{2N+2-2k}}-\frac{\beta_1^{2+l-k}}{r^{2N+2+2l-2k}})}{- 2\pi i t_1  (\beta_1-\beta_a)(\beta_1-\beta_b)}  d\beta_1 .
\end{align}
When $l+2-k < 0$, the term $G_{k,l}(\omega_2,-2)\sim (\beta_b^{l+2-k})/(r^{2l-2k+2N+2}\delta^2)= (\beta_a^{k-l-2})/(r^{2N-2}\delta^2) $. 
When $l-k+2 \geq 0$, the term $G_{k,l}(\omega_2,-2)\sim (\beta_a^{l+2-k})/(r^{2l-2k+2N+2}\delta^2)= (\beta_b^{k-l-2})/(r^{2N-2}\delta^2) $. In both of these two cases, we have $\abs{G_{k,l}(\omega_2,2)} \ll \abs{G_{k,l}(\omega_2,0)}$ .

At the last, we will discuss the right boundary effect for $q>0$, there is only one pole $\beta_a$ inside the cGBZ1, and it follows,
\begin{align}
     G_{k,l}(\omega_2,q) =(\frac{\delta}{t_2})^q \frac{(1-\frac{\beta_a^{2l}}{r^{2l}}-\frac{\beta_a^{2N+2-2k}}{r^{2N+2-2k}}+\frac{\beta_a^{2N+2+2l-2k}}{r^{2N+2+2l-2k}})\beta_a^{k-l+qN}}{-t_1(\beta_a-\beta_b)} (1-\beta_a^2/r^2)^q .
\end{align}
Due to the fact that $\abs{\beta_{a}}^{N}\delta \ll 1$, it will converge as $(\beta_{a}^{N}\delta)^q$ and the term $q=1$ is the leading term for $q>0$,
\begin{align}
     G_{k,l}(\omega_2,1) =\frac{\delta}{t_2} \frac{(1-\frac{\beta_a^{2l}}{r^{2l}}-\frac{\beta_a^{2N+2-2k}}{r^{2N+2-2k}}+\frac{\beta_a^{2N+2+2l-2k}}{r^{2N+2+2l-2k}})\beta_a^{k-l+N}}{-t_1(\beta_a-\beta_b)} (1-\beta_a^2/r^2) \sim \delta \beta_a^{k-l+N} .
\end{align}

Now we can conclude the leading term in the topological non-trivial regime. When $k \geq l$,  we have 
\begin{align}
    G_{k,l}(\omega_2,0) \sim \abs{\beta_a^{k-l}} \gg \abs{\beta_a^{k-l}\beta_a^N\delta}\sim \abs{\delta \beta_a^{k-l+N}} \sim  G_{k,l}(\omega_1,1) .
\end{align}
and the term $ \abs{G_{k,l}(\omega_2,0)} \gg \abs{G_{k,l}(\omega_2,-2)} $. We can find the Green's function will always be dominated by the term $G_{k,l}(\omega_2,0)$. When $k < l$, we need to compare the term $ G_{k,l}(\omega_2,1) \sim \delta\beta_a^{k-l+N}$ and $ G_{k,l}(\omega_2,0) \sim\beta_b^{k-l}$. For the response site close to the excitation site, i.e $k=l$, the term becomes
\begin{align}
    G_{k,l}(\omega_2,1) \sim \abs{\delta \beta_a^{k-l+N}} \sim \abs{\beta_a^N\delta} \ll 1 \sim \abs{\beta_b^{k-l}} \sim  G_{k,l}(\omega_2,0)  .
\end{align}
In this case the term $G_{k,l}(\omega_1,0)$ dominates. For the response site far way from the excitation site, i.e $k=1$ and $l=N$, the term reads
\begin{align}
     G_{k,l}(\omega_2,0) \sim \abs{\beta_b^{k-l}} \sim \abs{\beta_b^{-N}} \ll \abs{\beta_a^{-N}\beta_a^N\delta} \sim \abs{\delta\beta_a^{k-l+N}} \sim  G_{k,l}(\omega_2,1) .
\end{align}
In this case the term $G_{k,l}(\omega_2,1)$ dominates. Define a critical distance $N_0 = |k-l|$ at which the dominant contribution switches from $G_{k,l}(\omega_2,1)$ to $G_{k,l}(\omega_2,0)$ and it can be approximately calculated by taking $\abs{\delta\beta_a^{k-l+N}}=\abs{\beta_b^{k-l
}} $ with $N_0 \simeq \abs{(N\log \abs{\beta_a}+\log \delta)/(\log \abs{\beta_a}-\log\abs{\beta_b)}}$. The Green's function can be written as,
\begin{align}
    G_{k,l}(\omega_2)\simeq\left\{\begin{array}{l}
G_{k,l}(\omega_2,0) =  \oint_{\beta_1 \in \text{cGBZ1}} \frac{\beta_1^{k-l}-\frac{\beta_1^{k+l}}{r^{2l}}-(\frac{\beta_1^{2N+2-k-l}}{r^{2N+2-2k}}-\frac{\beta_1^{2N+2+l-k}}{r^{2N+2+2l-2k}})}{-2\pi i t_1(\beta_1-\beta_a)(\beta_1-\beta_b)}  d\beta_1 , k \geqslant l,l-k<N_0 \\
G_{k,l}(\omega_2,1) = \frac{\delta}{t_2}\oint_{\beta_1 \in \text{cGBZ1}} \frac{\beta_1^{k-l}-\frac{\beta_1^{k+l}}{r^{2l}}-(\frac{\beta_1^{2N+2-k-l}}{r^{2N+2-2k}}-\frac{\beta_1^{2N+2+l-k}}{r^{2N+2+2l-2k}})}{-2\pi it_1(\beta_1-\beta_a)(\beta_1-\beta_b)} \beta_1^{N}(1-\beta_1^2/r^2)d\beta_1, l-k \geq N_0 .
\end{array}\right.
\end{align}
Especially,  $G_{1,N}(\omega_2)\simeq-\delta (\beta_a-\beta_a^3/2)/(t_1t_2(\beta_a-\beta_b))$, which directly leads to the scale free suppression of the system. For the opposite direction response, $G_{N,1}(\omega_2) \simeq (\beta_a^{N-1}-\beta_a^{N+1}/r^2)/(t_1(\beta_a-\beta_b))$ which will lead to a size-dependent response amplification or suppression as $\beta_a^{N}$ depending on the position of $\omega_2$ relative to PBC spectrum.

\section{The anomalous scaling behavior}

\begin{figure}
    \centering
    \includegraphics[width=0.7\linewidth]{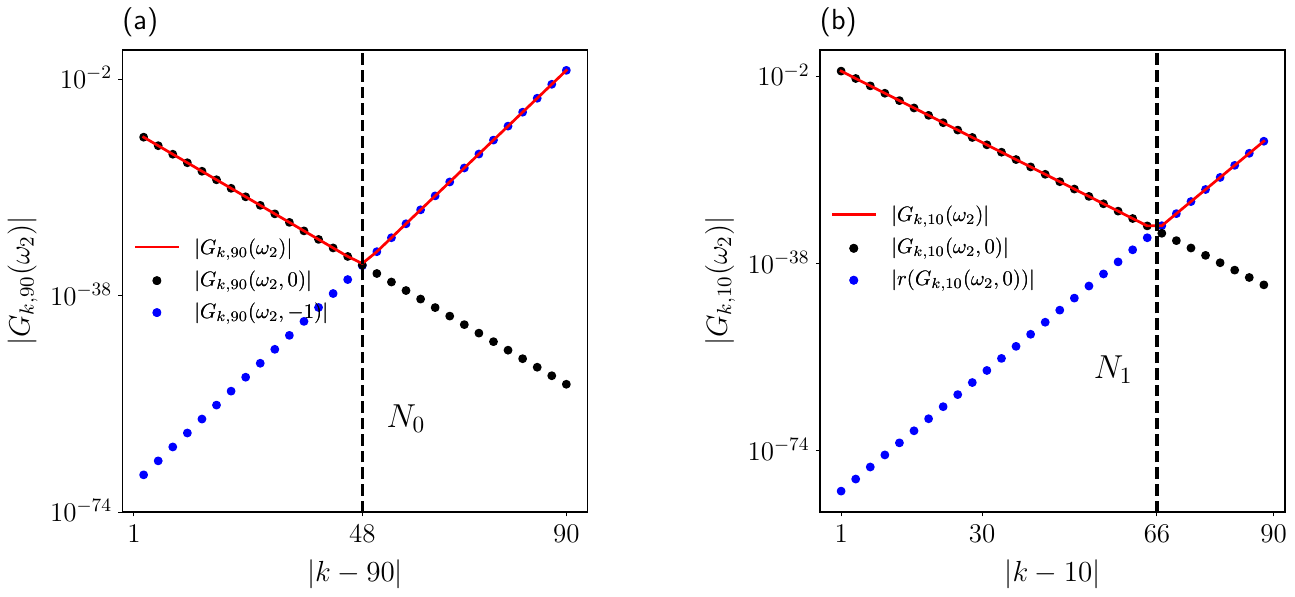}
    \caption{ The anomalous scaling behavior of the Green's function at $\omega_2 = 2.8i$ is shown in two directions. (a) For the response in the leftward direction , when $|k-90| < N_0 \simeq 48$, the analytic result $G_{k,90}(\omega_2,0)$ agrees with the numerical $G_{k,90}(\omega_2)$; when $|k-90| > N_0$, the analytic result $G_{k,90}(\omega_2,-1)$ agrees with the numerical result. (b) For the response in the opposite direction, the numerical result $G_{k,10}(\omega_2)$ agrees with the analytic result $G_{k,10}(\omega_2,0)$ before the critical distance $N_1 \simeq 66$, and with $r(G_{k,10}(\omega_2,0))$ beyond it. }
    \label{sup-fig1}
\end{figure}
At this section, we discuss the anomalous scaling behavior of Green's function which has been studied in article~\cite{d5zc-p1sk} concerning the critical non-Hermitian system. The regime for the driving frequency is in the  topological trivial regime with $W_{\text{cGBZ1},\omega_2}=0$ . The anomalous scaling behavior arises from the fact that, in different ranges of $\abs{k-l}$, the Green's function is dominated by different scaling rules.

For $k<l$, the anomalous scaling behavior comes from the competition between the $G_{k,l}(\omega_1,0)$ and $G_{k,l}(\omega_1,1)$ and the transition distance is determined by $N_0$. Such a behavior is illustrated in the last section. In the Fig.~\ref{sup-fig1}(a), we show that before the critical distance $N_0 \simeq 48$, the $G_{k,90}(\omega_1,0)$ is the leading term, and after the critical distance the $G_{k,90}(\omega_1,-1)$ is the leading term.

The anomalous scaling behavior for $k>l$ is caused by the correction term of $G_{k,l}(\omega_1,0)$. At the beginning,  we need to evaluate the influence of  an approximation taking  in the main text  that $(\delta\beta_1-t_1\beta_1^{N+1}) /(\delta\beta_2-t_1\beta_2^{N+1}) \simeq \beta_1^{N+1} /\beta_2^{N+1}$ and $(-t_1^{\ast}+\delta^{\ast}(\beta_1^{\ast})^{-N} )/(-t_1^{\ast}+\delta^{\ast}(\beta_2^{\ast})^{-N}) \simeq 1$ for solving $c_1$, $c_2$, $\tilde{c}_1$, $\tilde{c}_2$. In such an approximation, the effect of the term contained $\delta$ is ignored. We here change it to a more careful approximation,
\begin{align}
    \label{eq-sup1-correctioncoeff}c_1=\tilde{c}_1^{\ast}=1+\frac{\delta\beta_1^{-N}}{t_1}, \quad c_2=-\frac{\beta_1^{N+1}}{\beta_2^{N+1}}, \quad \tilde{c}_2=-1 .
\end{align}
This produces a correction to the term $G_{k,l}(\omega_1,0)$,
\begin{align}
    &G_{k,l}(\omega_2,0)+r(G_{k,l}(\omega_2,0))=\oint_{\beta_1 \in \text{cGBZ1}} \frac{\beta_1^{k-l}-\frac{\beta_1^{k+l}}{r^{2l}}-(\frac{\beta_1^{2N+2-k-l}}{r^{2N+2-2k}}-\frac{\beta_1^{2N+2+l-k}}{r^{2N+2+2l-2k}})}{-2\pi i t_1 (\beta_1-\beta_a)(\beta_1-\beta_b)(1+\delta\beta_1^{-N}/t_1)}  d\beta_1 \notag \\
    &+\oint_{\beta_1 \in \text{cGBZ1}} \frac{\frac{\delta\beta_1^{-N}}{t_1}(\beta_1^{k-l}-\frac{\beta_1^{k+l}}{r^{2l}})+\frac{\delta\beta_1^{-N}}{t_1}(\beta_1^{k-l}-\frac{\beta_1^{2N+2-k-l}}{r^{2N+2-2k}})}{-2\pi i t_1 (\beta_1-\beta_a)(\beta_1-\beta_b)(1+\delta\beta_1^{-N}/t_1)}  d\beta_1.
\end{align}
A correction term up to the first order of $\delta$ is
\begin{align}
    &r(G_{k,l}(\omega_2,0)) \simeq \oint_{\beta_1 \in \text{cGBZ1}} \frac{\delta(\beta_1^{k-l-N}-\frac{\beta_1^{N+2+l-k}}{r^{2N+2+2l-2k}})}{-2\pi i t_1^2 (\beta_1-\beta_a)(\beta_1-\beta_b)}  d\beta_1. 
 \end{align}
It can be calculated as,
\begin{align}
    r(G_{k,l}(\omega_2,0))      = \frac{\delta\beta_b^{k-l-N }(1-\frac{r^2}{\beta_b^2})}{t_1^2(\beta_b-\beta_a)} .
\end{align}
Compared with the result in the last section,
\begin{align}
     G_{k,l}(\omega_2,0) \simeq \frac{\beta_a^{k-l}-\frac{\beta_a^{k+l}}{r^{2l}}}{t_1(\beta_a-\beta_b)}.
\end{align}
Such an correction term can not be ignored for the frequency $\omega_2$ out of the PBC spectrum such that $\beta_a^N \ll \delta$. It can be seen that as $k-l$ close to $N$,
\begin{align}
    r(G_{k,l}(\omega_2,0)) \sim \delta \gg \beta_a^N \sim G_{k,l}(\omega_2,0).
\end{align}
In such a case, the correction term $r(G_{k,l}(\omega_2,0))$ is the leading term. As for $k-l$ close to $0$, we have 
\begin{align}
    r(G_{k,l}(\omega_2,0)) \sim \delta\beta_b^{-N} \ll 1 \sim G_{k,l}(\omega_2,0).
\end{align}
The transition length $N_1$ is given by $\abs{\delta\beta_b^{k-l-N}}=\abs{\beta_a^{k-l
}} $ with $N_1 \simeq \abs{(N\log \abs{\beta_b}-\log \abs{\delta})/(\log \abs{\beta_a}-\log\abs{\beta_b)}}$. In Fig.~\ref{sup-fig1}(b), before the critical distance $N_1 \simeq 66$, $G_{k,10}(\omega_2,0)$ is the leading term. After the critical distance, $r(G_{k,90}(\omega_2,0))$ becomes larger than $G_{k,10}(\omega_2,0)$ and dominates $G_{k,10}(\omega_2)$.